\documentstyle[12pt]{article}
\def\bea{\begin{eqnarray}}
\def\eea{\end{eqnarray}}
\def\nn{\nonumber}

\def\beq{\begin{equation}}
\def\eeq{\end{equation}}
\def\ba{\beq\new\begin{array}{c}}
\def\ea{\end{array}\eeq}
\def\be{\ba}
\def\ee{\ea}

\makeatletter
\newdimen\normalarrayskip              
\newdimen\minarrayskip                 
\normalarrayskip\baselineskip
\minarrayskip\jot
\newif\ifold             \oldtrue            \def\new{\oldfalse}
\def\arraymode{\ifold\relax\else\displaystyle\fi} 
\def\eqnumphantom{\phantom{(\theequation)}}     
\def\@arrayskip{\ifold\baselineskip\z@\lineskip\z@
     \else
     \baselineskip\minarrayskip\lineskip2\minarrayskip\fi}
\def\@arrayclassz{\ifcase \@lastchclass \@acolampacol \or
\@ampacol \or \or \or \@addamp \or
   \@acolampacol \or \@firstampfalse \@acol \fi
\edef\@preamble{\@preamble
  \ifcase \@chnum
     \hfil$\relax\arraymode\@sharp$\hfil
     \or $\relax\arraymode\@sharp$\hfil
     \or \hfil$\relax\arraymode\@sharp$\fi}}
\def\@array[#1]#2{\setbox\@arstrutbox=\hbox{\vrule
     height\arraystretch \ht\strutbox
     depth\arraystretch \dp\strutbox
     width\z@}\@mkpream{#2}\edef\@preamble{\halign
\noexpand\@halignto
\bgroup \tabskip\z@ \@arstrut \@preamble \tabskip\z@ \cr}%
\let\@startpbox\@@startpbox \let\@endpbox\@@endpbox
  \if #1t\vtop \else \if#1b\vbox \else \vcenter \fi\fi
  \bgroup \let\par\relax
  \let\@sharp##\let\protect\relax
  \@arrayskip\@preamble}
\def\eqnarray{\stepcounter{equation}%
              \let\@currentlabel=\theequation
              \global\@eqnswtrue
              \global\@eqcnt\z@
              \tabskip\@centering
              \let\\=\@eqncr
              $$%
 \halign to \displaywidth\bgroup
    \eqnumphantom\@eqnsel\hskip\@centering
    $\displaystyle \tabskip\z@ {##}$%
    \global\@eqcnt\@ne \hskip 2\arraycolsep
         $\displaystyle\arraymode{##}$\hfil
    \global\@eqcnt\tw@ \hskip 2\arraycolsep
         $\displaystyle\tabskip\z@{##}$\hfil
         \tabskip\@centering
    &{##}\tabskip\z@\cr}
\newfont{\hr}{msbm10}
\newfont{\ams}{msam10}

\begin{document}
\phantom.\hfill{ITEP/TH-71/97}\\
\phantom.\hfill{FIAN/TD-25/97}\\
\phantom.\hfill hepth/9712177
\vspace{0.5cm}
\begin{center}
{\LARGE \bf Covariance of WDVV Equations}
\vspace{0.5cm}

\setcounter{footnote}{1}
\def\thefootnote{\fnsymbol{footnote}}
{\Large A.Mironov\footnote{Theory
Department, Lebedev Physics Institute, Moscow
~117924, Russia; e-mail address: mironov@lpi.ac.ru}\footnote{ITEP,
Moscow 117259, Russia; e-mail address:
mironov@lpi.ac.ru, mironov@itep.ru}, A.Morozov\footnote{ITEP, Moscow
117259, Russia; e-mail address: morozov@vx.itep.ru}
}\\
\end{center}
\bigskip
\begin{quotation}
The (generalized)  WDVV equations for the prepotentials in
$2d$ topological and $4,5d$ Seiberg-Witten models
are covariant with respect to non-linear transformations,
described in terms of solutions of associated linear
problem. Both time-variables and the prepotential change
non-trivially, but period matrix (prepotential's second
derivatives) remains intact.
\end{quotation}


\section{Summary}
\setcounter{footnote}{0}
The WDVV (Witten-\-Dijkgraaf-\-Verlinde-\-Verlinde) equations
\cite{WDVV,D,KM}
form an overdefined set of non-linear equations for a
function (prepotential) of $r$ variables (times),
$F(t^i)$, $i=1,\ldots,r$. According to \cite{MMM},
they can be written in the form:
\be
F_i G^{-1} F_j = F_j G^{-1} F_i, \nn \\
G = \sum_{k=1}^r \eta^k F_k, \ \ \
\forall i,j = 1,\ldots,r \ \ {\rm and} \ \
\forall \eta^k(t)
\label{WDVV}
\ee
where $F_i$ are $r\times r$ matrices
$\displaystyle{(F_i)_{jk} = F_{,ijk} = \frac{\partial^3 F}
{\partial t^i\partial t^j\partial t^k}}$
and the "metric" matrix $G$ is an arbitrary linear
combination of $F_k$'s, with coefficients
$\eta^k(t)$ that can be time-dependent.\footnote{
The prepotential, as it arises in applications,
is naturally a homogeneous function of degree $2$,
but it depends on one extra variable $t^0$:
$$
{\cal F}(t^0,t^1,\ldots,t^r) = (t^0)^2F(t^i/t^0),
$$
see \cite{prep} for emerging the general theory.
As explained in \cite{BM}, the WDVV equations
(\ref{WDVV}) for $F(t^i)$ can be also rewritten
in terms of ${\cal F}(t^I)$:
$$
{\cal F}_I \hat{\cal G}^{-1} {\cal F}_J =
{\cal F}_J \hat{\cal G}^{-1} {\cal F}_I,  \ \ \
\forall I,J = 0,1,\ldots,r; \{ \eta^K(t)\}
$$
where this time ${\cal F}_I$ are $(r+1)\times(r+1)$ matrices
of the third derivatives of ${\cal F}$ and
$$
{\cal G} = \sum_{k=0}^r \eta^K {\cal F}_K, \ \ \
\hat{\cal G}^{-1} = (\det {\cal G}) {\cal G}^{-1}
$$
Note that homogeneity of ${\cal F}$ implies that
$t^0$-derivatives are expressed through those w.r.t. $t^i$, e.g.
$$
t^0{\cal F}_{,0ij}=-{\cal F}_{,ijk}t^k,\ \ \ \
t^0{\cal F}_{,00i}={\cal F}_{,ikl}t^kt^l,\ \ \ \
t^0{\cal F}_{,000}=-{\cal F}_{,klm}t^kt^lt^m \ \ \ \hbox{etc.}
$$
Thus, all "metrics" ${\cal G}$ are degenerate, but
$\hat{\cal G}^{-1}$ are non-degenerate.
Entire discussion of the present paper
allows one to put forward such reformulation in terms of
${\cal F}$, e.g. the Baker-Akhiezer vector-function
$\psi(t)$ should be just substituted by the explicitly
homogeneous (of degree 0) function $\psi(t^i/t^0)$.
The extra variable $t^0$ should not be mixed with the
distinguished "zero-time" associated with the constant metric
in quantum cohomology theory. Generically such a variable
does not exist (when it does, see comment 2.3 below,
we will identify it with $t^r$).
\label{f1}
}

The WDVV equations imply consistency of the following system
of differential equations \cite{M}:
\be
\left( F_{,ijk}\frac{\partial}{\partial t^l} -
       F_{,ijl}\frac{\partial}{\partial t^k} \right)\psi^j(t) = 0,
\ \ \ \forall i, j, k
\label{ls}
\ee
Contracting with the vector $\eta^l(t)$, one can also rewrite
it as
\be\label{*}
\frac{\partial \psi^i}{\partial t^k} = C^i_{jk} D\psi^j,
\ \ \ \forall i, j
\ee
where
\be\label{3}
C_k = G^{-1}F_k, \ \ \ G = \eta^lF_l,\ \ \ D = \eta^l\partial_l
\ee
(note that the matrices $C_k$ and the differential $D$ depend on
choice of $\{\eta^l(t)\}$, i.e. on choice of metric $G$)
and (\ref{WDVV}) can be rewritten as
\be\label{***}
\left[C_i,C_j\right]=0,
\ \ \ \forall i, j
\ee

The set of the WDVV equations (\ref{WDVV}) is invariant under
{\it linear} change of time variables with the prepotential
unchanged \cite{MMM}.
According to \cite{D} and especially to \cite{L},
there can exist also {\it non-linear} transformations
which preserve the WDVV structure, but they can change
the prepotential.
We shall show that such transformations are naturally
induced by solutions of the linear system (\ref{ls}):
\be
t^i \ \longrightarrow \ \tilde t^i = \psi^i(t), \nn \\
F(t) \ \longrightarrow \ \tilde F(\tilde t),
\label{tr}
\ee
so that the period matrix remains intact:
\be
F_{,ij} = \frac{\partial^2 F}{\partial t^i\partial t^j}
= \frac{\partial^2 \tilde F}{\partial \tilde t^i\partial \tilde t^j}
\equiv \tilde F_{,\hat i\hat j}
\label{pm}
\ee

\section{Comments \label{qh}}

{\bf 2.1.} As explained in \cite{M}, the linear system (\ref{ls})
has infinitely many solutions. "Original" time-variables
are among them: $\psi^i(t) = t^i$.

\noindent
{\bf 2.2.} Condition (\ref{pm}) guarantees that the transformation
(\ref{tr}) changes linear system (\ref{ls}) only by
(matrix) multiplicative factor, i.e. the set of solutions
$\{\psi^i(t)\}$ is invariant of (\ref{tr}). Among other
things this implies that repeated application of (\ref{tr})
does not provide new sets of time-variables.

\noindent
{\bf 2.3.} In the case of quantum cohomologies ($2d$ topological
models) \cite{WDVV,D,KM,L} there is a distinguished time-variable,
say, $t^r$, such that all $F_{,ijk}$ are independent of $t^r$:
\be
\frac{\partial}{\partial t^r} F_{ijk} = 0 \ \ \
\forall i,j,k = 1,\ldots, r
\ee
(equivalently, $\frac{\partial}{\partial t^i} F_{rjk} = 0$
$\forall i,j,k$).
Then (\ref{ls}) can be Fourier-transformed with respect to
$t^r$ and substituted by the system
\be
\frac{\partial}{\partial t^j} \hat\psi^i_z = zC^i_{jk}
\hat\psi^k_z,
\ \ \ \forall i, j
\ee
where
$\hat\psi^k_z(t^1,\ldots,t^{r-1}) =
\int \psi^k_z(t^1,\ldots,t^{r-1},t^r) e^{zt^r}dt^r$.
In this case the set of transformations (\ref{tr})
can be substituted by a family, labeled by a single variable $z$:
\be
t^i \ \longrightarrow \ \tilde t^i_z = \hat\psi^i_z(t)
\ee
In the limit $z \rightarrow 0$ and for the particular choice of the
metric, $\check G=F_r$, one obtains the particular
transformation
\be
{\partial \tilde t^i \over \partial t^j}
={\check C}^i_{jk}h^k, \ \ \ h^k = const,
\label{L}
\ee
discovered in \cite{L}. (Since ${\check C}_i=\partial_j \check C$, one can
also write ${\check C}_i={\check C}^i_kh^k$, ${\check
C}^i_k=\left(F_r^{-1}\right)^{il}F_{,lk}$.)

\noindent
{\bf 2.4.} Parametrization like (\ref{L}) can be used in generic situation
(\ref{tr}) as well (i.e. without distinguished $t^r$-variable and for
the whole family (\ref{tr})), only $h^k$ is not a constant,
but solution to
\be\label{10}
\left(\partial_j - DC_j\right)^i_kh^k = 0
\ee
($h^k = D\psi^k$ is always a solution, provided $\psi^k$
satisfies (\ref{*})).

\section{Infinitesimal variation of the WDVV equations}
In this section we look for the infinitesimal variation of the WDVV
equations, which preserve their shape. To this end, consider the small
variation of time-variables and the prepotential
\be\label{ch}
\tilde F(t)=F(t)+\epsilon f(t),\ \ \
t^i=\tilde t^i+\epsilon \xi^i(t)
\ee
This variation induces the variation of the third derivatives of the
prepotential
\be
{\tilde F}_{,\hat i\hat j\hat k}= F_{,ijk}+
\epsilon\left[\left\{f +F_{,l}\xi^l\right\}_{,ijk}-F_{,ijkl}\xi^l
\right]
\ee
The r.h.s. of this formula can be understood as matrix elements of the
matrix $F_j\left(I+\epsilon A_{j}\right)$ where we introduced
the new matrix $A$ with matrix elements defined by
\be
F_{,ijn}A^n_{jk}\equiv \left\{f +F_{,l}\xi^l\right\}_{,ijk}-F_{,ijkl}\xi^l
\ee
The form of the WDVV equations (\ref{WDVV}) is preserved by
the transformation (\ref{ch}) provided
\be
F_i\left(A_i-A_j\right)F_i^{-1}=F_k\left(A_k-A_j\right)F_k^{-1}
\ee
where the matrix $\left(A_i\right)^j_k\equiv A^j_{ik}$.
Solution to this equation is any constant ($i$-independent)
matrix $A$: $A_i= A$, $\forall i$. Therefore,
\be
F_{,ijn}A_k^n= \left\{f +F_{,l}\xi^l\right\}_{,ijk}-F_{,ijkl}\xi^l=
f_{,ijk}+\left(F_{,il}\xi^l_{,j}+ F_{,lj}\xi^l_{,i}+F_{,l}\xi^l
_{,ij}\right)
_{,k}+F_{,ijl}\xi^l_{,k}
\ee
The last term in the r.h.s. of this equation is of the desired form $F_j\times
A$, while it is hard to represent the remaining terms in this form.
Therefore, it is natural to request them vanish
\be\label{pminvi}
F_{,il}\xi^l_{,j}+ F_{,lj}\xi^l_{,i}+F_{,l}\xi^l_{,ij}=-f_{,ij}
\ee
This is nothing but the infinitesimal version of formula
(\ref{pm})! Therefore, we obtain the invariance of the period
matrix, eq.(\ref{pm}), as the condition of the WDVV system covariance.

Formula (\ref{pminvi}) allows one to find $f$ once the change of time
variables, i.e. functions $\xi^l(t)$, is known. However, one still needs to
find $\xi^i(t)$. In fact, they are restricted by the condition of
self-consistency of (\ref{pminvi}). Namely, the l.h.s. of  this formula
is the second derivative of something w.r.t.
$t^i$ and $t^j$ iff its third derivative w.r.t. $t^k$
is symmetric over all the three indices. This is equivalent to the
condition
\be
F_{,ijl}\xi^l_{,k}=F_{,kjl}\xi^l_{,i}
\ee
i.e. symmetricity under the
permutations of $i$ and $k$. This symmetry occurs
if $\xi^l$ have form
\be\label{fh}
\xi^l_{,i}=C^l_{ij}h^j
\ee
where $h^j$ are some new functions. Then, the symmetry is a simple corollary
of (\ref{WDVV}) and (\ref{*}).

Now, however, one should consider also restrictions on the functions $h^i$,
which can be derived from (\ref{fh}). Quite similarly to the previous step,
the r.h.s. of (\ref{fh}) is the derivative of something w.r.t.
$t^i$ iff its
derivative w.r.t. to $t^j$ is symmetric under
the permutation $i\leftrightarrow j$.
This condition is solved explicitly by the formula
\be\label{iter}
h^l_{,i}=D\left(C^l_{ij}h^j_{(1)}\right)
\ee
with some new functions $h^j_{(1)}$ that are also subject to the
consistency conditions. Iteratively repeating this procedure, one can express
$h^i_{(1)}$ through $h^j_{(2)}$ using formula (\ref{iter}) etc.

In fact, what we are doing in this way is the iterative procedure
($P$-exponential) for the following (matrix) equation
(the sign $\circ$
means the composition of operators)
\be
h^l_{,i}=D\left(C^l_{ij}h^j\right), \ \ \ \hbox{i.e. }\
\left(\partial_i - D\circ C_i\right) h=0
\ee
Therefore, we come to formula (\ref{10}). In the next section we check
that this equation provides also the non-infinitesimal transformation of
time-variables. This is not surprising, since it is linear.

\section{Proof of covariance}
Now we investigate the non-infinitesimal transformations of time-variables
$t^i\to\psi^i$ given by the formula
\be\label{p1}
{\partial\psi^l\over\partial t^i}=\left(C^l_{ij}h^j\right)
\ee
where functions $h^i$ satisfy the equation (\ref{10}).
The proof is a combination of
reasoning from \cite{L,M}.

First of all, let us
check that the differential operators ${\cal D}_i\equiv \partial_i
-D\circ C_i$ in eq.(\ref{10})
induce a self-consistent system, i.e. they are commuting.
Indeed, one can use the identity
\be\label{identity}
\partial_j C_i-\partial _i C_j=C_j\left(DC_i\right)-C_i\left(DC_j\right),
\ee
i.e.
\be\label{identity2}
D\left(\partial_i C_i-\partial _iC_j\right)=\left[DC_j,DC_i\right]
+C_jD^2C_i-C_iD^2C_j
\ee
In order to get (\ref{identity}), we used the WDVV equations and the
definitions (\ref{*}) and the identity $\partial_l F_{,ijk}=\partial_i
F_{ljk}$.
Then, one can easily see that
\be
\left[D_i,D_j\right]=D\left(\partial_i C_i-\partial _iC_j\right)
\left[DC_i,DC_j\right] +C_iD^2C_j-C_jD^2C_i +\\+
\left(C_iDC_j-DC_jC_i+DC_iC_j-C_jDC_i
\right)D=\\=D\left(\partial_i C_i-\partial _iC_j\right)
\left[DC_i,DC_j\right] +C_iD^2C_j-C_jD^2C_i +
D\left(\left[C_i,C_j\right]\right)D=0
\ee
because of (\ref{identity2}) and (\ref{***}).

Now we should check the consistency of the change of
time-variables (\ref{p1}), i.e. that
the r.h.s. of formula (\ref{p1})
can be presented as the derivative of something
w.r.t.
$t^i$. To this end, as usual, we should check the symmetricity of the
derivative of (\ref{p1}) w.r.t.
$t^k$ under the permutation $i\leftrightarrow
k$. This is equivalent to the condition that $\psi^i_{,jk}$ is symmetric
under the permutation of indices $j$ and $k$. Indeed,
\be
\psi^i_{,jk}=C_j\partial_k h+\left(\partial_l C_j\right)h=C_jC_kDh+
\left(C_jDC_k+\partial_lC_j\right)h
\ee
This expression is really symmetric because of (\ref{***}) and
(\ref{identity}).

The next step of our proof is to check the invariance of the period matrix
(\ref{pm}), i.e. existence of a function $\tilde F$ such that (\ref{pm})
is
fulfilled. In other words, one needs to check that $F_{,ij}$ can be presented
as a second derivative w.r.t. the time-variables $\{\psi^i\}$. Therefore, one
should again take the derivative of $F_{,ij}$ w.r.t. $\psi^k$ and check its
symmetricity:
\be
{\partial F_{,ij}\over\partial\psi^k} =F_{,ijl}{\partial t^l\over
\partial\psi^k}
\equiv \left(F_j U\right)_{ik}
\ee
where $U$ is the matrix with elements $\displaystyle{
\left(U\right)^l_k={\partial t^l\over \partial\psi^k}}$.
Now, instead of checking the symmetricity of the matrix $F_jU$, we check the
symmetricity of the inverse matrix $U^{-1}F_j^{-1}$. Its
matrix elements are (see (\ref{p1}))
\be
C^i_{lm}\left(F_j^{-1}\right)^{lk}h^m=\left(G^{-1}F_mF^{-1}_j\right)^{ik}h^m
\ee
This matrix is indeed symmetric, since every one of matrices $F_i$ is
symmetric and because of the WDVV equations.

Thus, we proved that our construction is consistent. It remains to
check that the WDVV equations are
covariant under the change $F_i\to\tilde F_i$, $t^i\to \psi^i$. Indeed,
\be
\tilde F_i\tilde G^{-1}\tilde F_j=F_iUU^{-1}G^{-1}F_jU=
F_jG^{-1}F_iU=\tilde F_j\tilde G^{-1}\tilde F_i
\ee
This completes the proof.

Note that, from (\ref{p1}) and (\ref{10}), it follows that the new
time-variables $\psi^k$ are solutions to the equation (cf. (\ref{3}))
\be
\left(\partial_i-C_iD\right)\psi=0
\ee
provided $h^k=D\psi^k$. Therefore, they solve the linear problem for the WDVV
equations.

\section{Conclusion and acknowledgments}

To conclude, we described a set of non-trivial non-linear
transformations which preserve the structure of the WDVV
equations (\ref{WDVV}). The consideration above does not
{\it prove} that {\it all} such transformations are of the
form (\ref{tr}), (\ref{pm}), and even in this sense
the story is incomplete.
Still (\ref{tr}) is already unexpectedly(?) large,
because (\ref{WDVV}) is an {\it overdefined} system and
it could seem to be {\it rigid}, if has any solutions
at all.

Even more obscure are the {\it origins and implications}
of this covariance.
Two remarks deserve to be made.

First, we see that the "period matrix" $F_{,ij}$
appears "more rigid" than the prepotential itself:
according to (\ref{pm}) it does not change under
(\ref{tr}).\footnote{In terms of footnote \ref{f1}, $f_{,IJ}$
is a homogeneous function of degree $0$, which one can naturally
expect to be more "stable" than ${\cal F}$, which is a
homogeneous function of degree $2$.}
Note that, in the framework of Seiberg-Witten theory \cite{SW},
$F_{,ij}$, not the prepotential $F$ itself, has the
direct physical meaning (it describes coupling constants
of the low-energy abelian effective theory).
Thus, in certain sense, the transformations
(\ref{tr}), (\ref{pm}) leave "physics" invariant,
as one could wish.

Second, an essential ingredient of the Seiberg-Witten theory
is its hidden integrable structure \cite{int,int_rev}
which is also relevant for consistency with the
brane theory \cite{br}. In this context, the role of time-variables
$t^i$ is played by the periods $a^i$ of the presymplectic
$1$-form (of some $0+1$-dimensional quantum-mechanical system),
$t^i = a^i \equiv \oint_{A_i} dS$ on the family of spectral curves.
It is unclear if the transformations (\ref{tr}) can be lifted
to some deformation of this structure.
Note that ref.\cite{Giv} suggests that the problem of integral
representation for $\psi^i(t)$ (defined as a solution of the linear
system (\ref{ls})) is related to that of the mirror-like maps.

\bigskip

We acknowledge discussions with A.Gorsky, A.Marshakov and especially
A.Losev.
This work was partly supported by the grants
RFFI 96-02-19085, INTAS 96-482 (A.Mir.)
and RFFI 96-15-96939 (A.Mor).

\end{document}